\documentclass[preprint,prd,groupedaddress,footinbib,showpacs,rotate]{revtex4}

\usepackage{graphicx}
\usepackage{dcolumn}
\usepackage{psfig}
\def \R{\rm l\!R\,}

\begin{document}

\title{Anti-de Sitter boundary in Poincar\'e coordinates}
\author{C.  A.  Ball\'on Bayona}
\email{ballon@if.ufrj.br}
\affiliation{Instituto de F\'{\i}sica,
Universidade Federal do Rio de Janeiro, Caixa Postal 68528, RJ 21941-972 -- Brazil}
\author{Nelson R. F. Braga}
\email{braga@if.ufrj.br}
\affiliation{Instituto de F\'{\i}sica,
Universidade Federal do Rio de Janeiro, Caixa Postal 68528, RJ 21941-972 -- Brazil}


\begin{abstract}
We build up a detailed description of the space-time boundary of a Poincar\'e patch of anti-de
Sitter ($AdS$) space. We map the Poincar\'e $AdS$ boundary to the
global coordinate chart and show why this boundary is not
equivalent to the global $AdS$ boundary. The Poincar\'e $AdS$
boundary is shown to contain points of the bulk of the entire
$AdS$ space. The Euclidean $AdS$ space is also discussed. In this
case one can define a semi-global chart that divides the $AdS$
space in the same way as the corresponding Euclidean Poincar\'e
chart.

\end{abstract}


\maketitle

\section{ Introduction }

Some years ago, Maldacena found very important dualities between
string theories and conformal field theories (CFT) known
collectively as $AdS$/CFT correspondence
\cite{Maldacena:1997re,Petersen:1999zh,Aharony,D'Hoker:2002aw}.
A particularly important example relates string theory in a ten
dimensional curved space with super-conformal Yang Mills theory in
four dimensional flat space. The starting point for finding this
duality was a solution of low energy string theory containing
D3-brane charge. The corresponding metric, studied in \cite{HS}, is
similar to that of a black hole. The near horizon limit of this
D3-brane metric corresponds to the direct product of a Poincar\'e
chart of $AdS_5\,$ space and a compact $S^5 $ space. String theory
in this ten dimensional space, referred for short, as $AdS_5\times
S^5\,$ is dual to the four dimensional CFT. As we will see in
detail, a Poincar\'e chart represents only a region of the entire
$AdS$ space. Since this distinction will be important here, we
will call this region as: Poincar\'e $AdS$ space.

Quantization of fields in global $AdS$ space requires the introduction of a
boundary, at spatial infinity, where vanishing flux conditions can
be imposed. Otherwise particles could enter or leave the space in
finite times \cite{Avis:1977yn,Breitenlohner:1982bm}. 
This boundary has the topology of $S^1\times S^{n-1}\,$. 
The boundary of a Poincar\'e $AdS$ space is a different region. 
This boundary assumes a special role in
the $AdS$/CFT correspondence.
It was shown in \cite{m3,Witten:1998qj} that correlation functions
for the CFT can be calculated in terms of string theory in such a
way that the CFT is interpreted as living on the boundary of the
Poincar\'e $AdS$ space. So, we can think of the $AdS$/CFT
correspondence as a realization of the holographic
principle \cite{HOL1,HOL2} since it relates a theory in a space
with gravity to a theory on its boundary.
It is important to remark that although the $AdS$/CFT correspondence 
was originally formulated for a Poincar\'e AdS space, it also holds 
for $AdS$ space in global coordinates\cite{Aharony}.
Previous related discussions of the $AdS$ boundary
can be found in \cite{Witten2,Boschi-Filho:2001gr}.
Field quantization in Poincar\'e $AdS$ space  were discussed 
for example in \cite{Balasubramanian:1998sn,Boschi-Filho:2000vd}.

The AdS /CFT prescription for relating bulk and boundary theories is \cite{Witten:1998qj}

\begin{equation}
\label{Gen}
exp \Big(  - S (\phi ) \, \Big)  \,=\, \langle exp \int \phi_0 {\cal O}\,\rangle
\end{equation} 

\noindent where $S (\phi )$ is the on shell action for the bulk field $\phi$ and $\phi_0$ is proportional to the boundary value of $\phi$. This $\phi_0$ plays the role of an external source for the boundary conformal operators ${\cal O}$. Using this equation one can calculate the conformal correlation functions.

 In order to calculate correlation functions one must choose a coordinate system for the AdS space. 
This choice has non trivial consequences. As discussed in detail in 
ref \cite{Emparan:1999pm}, for the Euclidean case, different coordinate charts of AdS space can lead to different boundaries. 
Depending on the choice of metric, spaces that are locally equivalent to $AdS_{n+1}\,$  can have boundaries like, for example,  $ \R^n \,,\, S^1\times S^{n-1} \,, \,S^1 \times \R^{n-1}\,, 
\,S^1 \times H^{n-1} \,$. 
The different metrics representing AdS space are related by diffeomorphisms (coordinate transformations) plus  the inclusion of points at infinity and some identifications. 
From equation (\ref{Gen}), where the on shell action is a boundary term, we see that the dual conformal field theories are defined in different spaces, depending on the choice of the AdS metric. 
These theories will in general have different energies
as shown in \cite{Emparan:1999pm}. The difference in mass in global and Poincar\'e coordinates is
also  discussed for example in \cite{Balasubramanian:1999re}.

Another example of the non trivial role of the choice of AdS coordinate systems
is the case of the stress tensor. For asymptotically AdS spaces one can use the AdS/CFT correspondence to relate a gravitational stress tensor to the expectation value of the corresponding conformal 
stress tensor\cite{Balasubramanian:1999re}.  This boundary tensor diverges but can be regularized adding 
counterterms to the action. This subtraction procedure depends on the geometry of the boundary.
Different metrics will lead to different expectation values of the conformal stress tensor
(see also, for example,  \cite{Astefanesei:2004kn}).  
The choice of coordinate system can also change the stability of 
the AdS space when one consider AdS/CFT at finite temperature\cite{Witten:1998zw}. 
For global coordinates, there is phase transition to a black hole geometry at some critical temperature.
For Poincar\'e coordinates there is no thermal phase transition.

In this work, we will begin describing how the Poincar\'e $AdS$ space arises from a
non-trivial change of coordinates for the entire $AdS$ space. Then
we will present a division of the Poincar\'e $AdS$ boundary into
regions for which we find  well defined mappings to global
coordinates. This way the Poincar\'e $AdS$ boundary will be
completely described in the global chart.

\section{Global and Poincar\'e coordinates for the Anti-de Sitter
space}

Anti-de-Sitter space in $n+1$ dimensions can be  represented
 as a hyperboloid of radius $R$

\begin{equation}
{X_{0}}^2+{X_{n+1}}^2-\sum_{i=1}^{n}{{X_{i}}^2}= R^2 \, ,
\label{emb}
\end{equation}

\noindent embedded in an n+2 dimensional flat  space with metric

\begin{equation}
ds^2= -dX_{0}^2-dX_{n+1}^2+\sum_{i=1}^{n}{{dX_{i}}^2} \, .
\label{met}
\end{equation}

\noindent The coordinates $X_{a}$ \, for $a=0,..,n+1$ are known
as embedding coordinates. We can solve eq. (\ref{emb}) by introducing
the following relations

\begin{eqnarray}
X_{0}&=&R\sec{\rho}\cos{\tau} \, ,
\nonumber\\
X_{i}&=&R\,\tan{\rho}\,\Omega_{i} \, , \hskip1cm i=1,..,n \,  ,
\nonumber\\
X_{n+1}&=&R\sec{\rho}\sin{\tau}  \label{glob}\, ,
\end{eqnarray}

\noindent where $0\le\rho<\pi/2\,$,  $ \,-\pi < \tau\le\pi\,$
and $ -1 \le  \Omega_{i} \le 1\,$.
Coordinates $\rho\,,\tau \, $ and  $\Omega_{i}$ represent all the
hyperboloid and are called global coordinates.
The $\Omega_{i}$ are not independent. They must satisfy
the constraint $ \,\sum_{i=1}^{n}{{\Omega_{i}}^2}=1\,$.
The $AdS$ metric in terms of these coordinates reads
\begin{equation}
ds^2=\frac{R^2}{\cos^2\rho}(-d\tau^2+d\rho^2+\sin^2\rho
\sum_{i=1}^{n}{(d\Omega_{i})^2} )\, .
\end{equation}

In these global coordinates the $AdS$ boundary is the hypersurface $\rho = \pi/2\,$.
This corresponds, in embedding coordinates, to the spatial infinity.
So, we can compactify the $AdS$ space by changing the
range of the radial coordinate $\rho\,$ to: $0 \le \rho \le \pi/2$.

\bigskip
On the other hand, the Poincar\'e coordinate system can be
introduced by first defining the light cone coordinates :
\begin{eqnarray}
u &\equiv& \frac{(X_{0}-X_{n})}{R^2} \, ,
\nonumber \\
v &\equiv& \frac{(X_{0}+X_{n})}{R^2} \, . \label{poinc1}
\end{eqnarray}

\noindent This change of coordinates will make it possible to
absorb the time-like coordinates $X_{0}$. Redefining the other
coordinates as

\begin{eqnarray}
x^{i}&\equiv& \frac{X_{i}}{R u} \, \, \, \,\,\,\,\,\,\hskip 1cm
\hbox{(space-like)}\, ,
\nonumber  \\
t &\equiv& \frac{X_{n+1}}{R u} \, \, \, \,  \hskip 1cm \hbox{(time-like)}
\label{poinc2}\,,
\end{eqnarray}

\noindent equation (\ref{emb}) for the hyperboloid
takes the form
\begin{equation}
R^4uv + R^2u^2(t^2-{\overline{x}}^2) = R^2 \, ,
\end{equation}

\noindent where ${\overline{x}}^2\equiv \sum_{i=1}^{n-1}(x^{i})^2$. From this equation
we can express $v$ in terms of $ u\,,t\,,$ and $x^i \,$
and find from  (\ref{poinc1}) and (\ref{poinc2})
\begin{eqnarray}
X_{0} &=& \frac{1}{2u} (1+u^2(R^2+{\overline{x}}^2-t^2)) \, ,
\nonumber \\
X_{n} &=& \frac{1}{2u} (1+u^2(-R^2+{\overline{x}}^2-t^2)) \, ,
\nonumber\\
X_{i} &=& R u x^{i} \hskip1cm i=1,..,n-1 \, ,
\nonumber \\
X_{n+1} &=&  R u t \, .
\end{eqnarray}

\noindent  It is useful to change to the coordinate
$z\equiv\frac{1}{u}$. This way the Poincar\'e coordinates $z,
\overline{x} , t $ are defined by the following relations
\begin{eqnarray}
X_{0} &=& \frac{1}{2z}(z^2+R^2+{\overline{x}}^2-t^2) \, ,
\nonumber\\
X_{i} &=& \frac{R x^i}{z} \,  , \hskip1cm i=1,..,n-1 \, ,
\nonumber\\
X_{n} &=& \frac{1}{2z}(z^2-R^2+{\overline{x}}^2-t^2) \, ,
\nonumber\\
X_{n+1} &=& \frac{R t}{z} \, .\label{Poinc}
\end{eqnarray}

The $AdS$ metric in terms of these coordinates takes the nice form

\begin{equation}
ds^2=\frac{R^2}{z^2}(dz^2+(d\overline{x})^2-dt^2) \, .
\end{equation}

\noindent The coordinate $z$ behaves as a radial coordinate and
divide the $AdS$ space in two regions. Noting from (\ref{poinc1})
that

\begin{equation}
\label{Z}
\frac{1}{z} \,=\, u = \frac{(X_{0}-X_{n})}{R^2}\, ,
\end{equation}

\noindent we conclude that we have two different Poincar\'e
charts. The first chart is the region $z>0$ , that means $X_{0}>
X_{n}$ and corresponds to one half of the hyperboloid. In global
coordinates this region can be obtained by imposing the condition
$\cos{\tau}>\Omega_{n}\sin{\rho}$. The other half of the
hyperboloid $X_{0}< X_{n}$ corresponds to the second Poincar\'e
chart, i.e, the region $z < 0$ \, ( in global coordinates the
condition is $\cos{\tau}<\Omega_{n}\sin{\rho}$). The Poincar\'e
$AdS$ space is the region of the entire $AdS$ corresponding to one of
these two charts (the $z>0$ is usually chosen). Equation (\ref{Z}) shows us that
the region $z=0$ does not belong to the $AdS$ space. As we will see, it is part of
the $AdS$ boundary.

\begin{figure}[!h]
\centerline{\psfig{figure=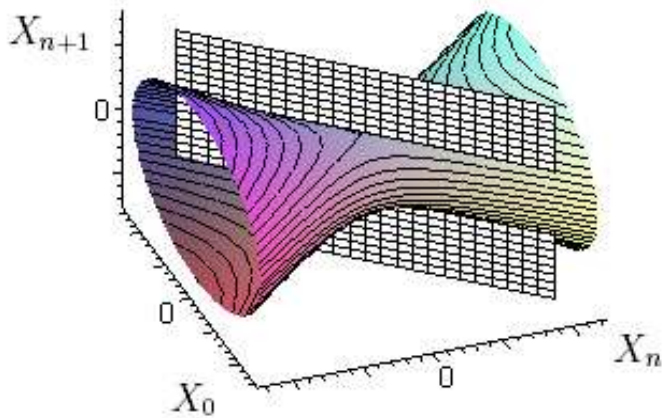,width=10cm}}
\vspace{-1cm}\caption{$AdS$ space being cut by the hyperplane
$X_{0}=X_{n}$ (coordinates $X_{i}$ fixed for $i=1,..,n-1$).
\label{Fig1}} \
\end{figure}

The hyperplane $X_{0}\,=\, X_{n}$, that cuts the
$AdS$ space into the two Poincar\'e charts, is not contained in these
charts but rather corresponds to the limits $z \to \pm \infty $ (or $ u=0$).
We can see in Figure \ref{Fig1} how the hyperplane $X_{0}\,=\,
X_{n}$ cuts the entire $AdS$ space. In this figure we consider the coordinates
$X_{i}\,$ ($i=1,..,n-1$) fixed. In this case the region $X_{0}\,=\,X_{n}$
is a plane that intersects the hyperboloid in two straight lines.

There are points of the hyperboloid contained in the cutting
hyperplane. From (\ref{Poinc}) we note that these points require
also the condition $t\to\pm \infty$ in order to satisfy
(\ref{emb}). In global coordinates these points must satisfy the
condition

\begin{equation}
\label{CUTPLANE} \cos{\tau}\,=\,\Omega_{n}\sin{\rho} \, ,
\end{equation}

\noindent where $0 \le \rho<\pi/2$. It is interesting to analyze
 the limit $\rho\to\pi/2$, which means: going to
spatial infinity. Equation (\ref{CUTPLANE}) reduces to
$\cos{\tau}=\Omega_{n}$ in this limit. This condition is very
special because it can be obtained also for z finite.
This happens because $\,\cos{\tau}-\Omega_{n}\,\to \,0\,$
when $\rho\to\pi/2$ with $ \,0\,< \vert z \vert \,< \infty \,$.
All these points will be studied
in detail in the next section.

\vspace{-.3cm}

\section{ Poincar\'e $AdS$ boundary in global coordinates}

In order to map the Poincar\'e $AdS$ boundary to global coordinates
we first obtain relations between these two coordinates systems.
From eq. (\ref{glob}) and eq. (\ref{Poinc}) one can find the
relations

\begin{eqnarray}
\sec{\rho} &=& \frac{1}{2R
|z|}\sqrt{(z^2+R^2+{\overline{x}}^2-t^2)^2
+ (2R t )^2 } \, , \label{rho} \\
\cos{\tau}&=&sgn(z)
\frac{z^2+R^2+{\overline{x}}^2-t^2}{\sqrt{(z^2+R^2+{\overline{x}}^2-t^2)^2+(2R
t)^2}} \, \, , \, \, \, \hbox{where} \, \,  sgn(z)\equiv
\frac{|z|}{z}  \, ,
\label{tau1} \\
\sin{\tau}&=&sgn(z)\frac{2R
t}{\sqrt{(z^2+R^2+{\overline{x}}^2-t^2)^2+(2R
t)^2}} \, , \label{tau2} \\
 |\overline{\Omega}| \equiv
\sqrt{\sum_{i=1}^{n-1}(\Omega_{i}^2)}&=&\frac{2R
|\overline{x}|}{\sqrt{(z^2+R^2+{\overline{x}}^2-t^2)^2 + (2R
t)^2-(2R z)^2}} \, \, , \, \hbox{where} \, |\overline{x}| \equiv
\sqrt{\overline{x}^2} \, \, , \label{omegai} \\
\Omega_{n}&=&sgn(z)\frac{z^2-R^2+{\overline{x}}^2-t^2}
{\sqrt{(z^2+R^2+{\overline{x}}^2-t^2)^2 + (2R t)^2 - (2R z )^2}}
\, . \label{omegan}
\end{eqnarray}

\noindent For instance, the first relation can be obtained by
adding the squares of the first and last relations in eqs.
(\ref{glob}) and (\ref{Poinc}). These relations are valid for $ z
> 0$ and $z<0$. For our
discussion we will consider just the Poincar\'e chart $z > 0$.

We present in table I a possible division of the Poincar\'e $AdS$ boundary into
regions for which  the coordinate transformations
(\ref{rho}-\ref{omegan}) lead us to well defined points in global
coordinates.

\begin{table} \centering
\begin{tabular}{ | c | c | c | c | c | c | c | c |}
\hline Region & z $\,\,$
 &  t $\,\,$   &
$ |\overline{x}|$  $\,\,$ & $\rho$ & $\tau$ & $|\overline{\Omega}|$ & $\Omega_{n}$\\
\hline
 I \,\,     & $z = 0^{+}$  & finite & finite & $\pi/2$ & $\tau^{I}$ &
$|\overline{\Omega}|^{I}$ & $(\Omega_{n})^{I}$ \\
 \hline
 II  & finite   &  $t \to \pm \infty$ & finite & $\pi/2$ & $\pm \pi$ & 0 & -1\\
 \hline
 III  & finite   &  finite & $|\overline{x}|\to  \infty$ & $\pi/2$ & 0 & 0 & 1\\
\hline
IV & finite    & $t \to \pm \infty$  & $|\overline{x}|\to  \infty$  ,
$|\overline{x}|<<|t|$& $\pi/2$ & $\pm \pi$ & 0 & -1  \\
\hline
 V & finite   & $t \to \pm \infty $  & $ |\overline{x}|\to  \infty$  ,
$|\overline{x}|>>|t|$ & $\pi/2$ & 0 & 0 & 1  \\
\hline
 VI & finite    & $t \to  \infty $  & $|\overline{x}|\to  \infty$  ,
$ \overline{x}^2=t^2+ \alpha|\overline{x}|$ &
$\pi/2$ & $\tau^{VI}$ & $|\overline{\Omega}|^{VI}$ &
$(\Omega_{n})^{VI}$  \\
\hline
 VII & finite    & $t \to - \infty $  & $|\overline{x}|\to  \infty$  ,
$\overline{x}^2=t^2+ \alpha|\overline{x}|$ & $\pi/2$ & $\tau^{VII}$ &
$|\overline{\Omega}|^{VII}$ &
$(\Omega_{n})^{VII}$  \\
\hline
 VIII    & $z \to \infty$  & finite & finite & $\pi/2$ & 0 & 0 & 1  \\
 \hline
 IX  & $z \to \infty$   &  $t \to \pm \infty$  ,
$z<<|t|$ & finite  & $\pi/2$ & $\pm \pi$ & 0 & -1 \\
\hline  X  & $z \to \infty$   &  $t \to \pm \infty$  ,
$z>>|t|$  & finite & $\pi/2$ & 0 & 0 & 1  \\
\hline  XI  & $z \to \infty$   &  $t \to \infty$  ,
$z^2=t^2+\alpha z$ \,   & finite & $\rho^{XI}$ & $\tau^{XI}$ &
$|\overline{\Omega}|^{XI}$ & $(\Omega_{n})^{XI}$
\\
\hline  XII  & $z \to \infty$   &  $t \to - \infty$  ,
$z^2=t^2+\alpha z$ \,    & finite & $\rho^{XII}$ & $\tau^{XII}$ &
$|\overline{\Omega}|^{XII}$ & $(\Omega_{n})^{XII}$
\\
\hline  XIII  & $z \to \infty$   &  finite
  & $|\overline{x}|\to  \infty$ & $\pi/2$ & 0 & 0 & 1
\\
\hline  XIV  & $z \to \infty$   & $t \to \pm \infty$  ,  $z^2+
\overline{x}^2 << t^2$
  & $|\overline{x}|\to  \infty$ & $\pi/2$ & $\pm \pi$ & 0 & -1
\\
\hline  XV  & $z \rightarrow \infty$   &  $t \to \pm \infty$  ,  $z^2+
\overline{x}^2 >> t^2$ & $|\overline{x}|\to  \infty$ & $\pi/2$ & 0
& 0 & 1
   \,
\\
\hline  XVI  & $z \to \infty$   &  $t \to \infty$  ,    & $|\overline{x}|\to \infty$  ,
 $\overline{x}^2=\beta^2 z^2$ \,
   \, & $\rho^{XVI}$ & $\tau^{XVI}$ &
$|\overline{\Omega}|^{XVI}$ & $(\Omega_{n})^{XVI}$
\\
 & & $z^2+
\overline{x}^2 = t^2 + \alpha z$ & & & & & \\
\hline  XVII  & $z \to \infty$   &  $t \to - \infty$  ,  & $|\overline{x}|\to \infty$  ,
$\overline{x}^2=\beta^2 z^2$ \,& $\rho^{XVII}$ & $\tau^{XVII}$ &
$|\overline{\Omega}|^{XVII}$ & $(\Omega_{n})^{XVII}$
\\
& & $z^2+
\overline{x}^2 = t^2 + \alpha z$ & & & & &
\\
\hline
\end{tabular}
\parbox{5in}{\caption{Poincar\'e $AdS$ boundary
 divided into different regions
corresponding to well defined points in global coordinates. The
parameters $\alpha$ and $\beta$ are finite with $\beta \ne 0$. \sl
}}
\end{table}

In this table we use the following definitions :

\begin{itemize}
\item In region I , we have that $\rho^{I}=\pi/2$. The coordinates
$\tau^{I},|\overline{\Omega}|^{I}$ and $(\Omega_{n})^{I}$ are
given by the relations

\begin{eqnarray}
\cos{\tau^{I}}&=&\frac{R^2+{\overline{x}}^2-t^2}{\sqrt{(R^2+{\overline{x}}^2-t^2)^2+(2R
t)^2}} \,  , \nonumber \\
 \sin{\tau^{I}}&=&\frac{2R
t}{\sqrt{(R^2+{\overline{x}}^2-t^2)^2+(2R t)^2}} \, , \nonumber \\
|\overline{\Omega}|^{I}&=& \frac{2R
|\overline{x}|}{\sqrt{(R^2+{\overline{x}}^2-t^2)^2 + (2R t)^2}} \,
, \nonumber \\
(\Omega_{n})^{I}&=&\frac{-R^2+{\overline{x}}^2-t^2}{\sqrt{(R^2+{\overline{x}}^2-t^2)^2
+ (2R t)^2 }} \, .
\end{eqnarray}

\noindent  This region corresponds to the hyperplane $z=0$ . It is
 interpreted as the Minkowski space-time and is part of the global $AdS$ boundary .

\item Regions VI and VII satisfy the condition $\cos{\tau}=\Omega_{n}$
 for $z$ finite (this condition was discussed in the previous section).
 These regions belong to the global $AdS$ boundary because $\rho=\pi/2$.

\noindent  The points of region VI are defined by the
following relations :

 \begin{eqnarray}
\cos{\tau^{VI}}&=&(\Omega_{n})^{VI}=\left\{
\begin{array}{ll}\pm
\frac{1}{\sqrt{1+(\frac{2R}{\alpha})^2}} & , \mbox{for
$sgn(\alpha)=\pm 1$ }\\
        0 & ,\mbox{for $\alpha=0$} \end{array}
         \right. \, ,  \, \,
0\le\tau^{VI}\le\pi \, , \, \, \nonumber
\\
|\overline{\Omega}|^{VI}&=&
\frac{1}{\sqrt{1+(\frac{\alpha}{2R})^2}} \, \, \, .
 \end{eqnarray}

\noindent Region VII has the same points as in VI with the difference
that $-\pi\le\tau^{VII}\le 0$ \, .

\item Regions XI, XII, XVI and XVII  correspond to special
cases of the limit $z\to\infty$  , $t\to\pm\infty$. These regions
belong to the cutting hyperplane and the hyperboloid. The points
of region XI are defined by

\begin{eqnarray}
\sec{\rho^{XI}}&=&\sqrt{1+(\frac{\alpha}{2R})^2} \,\,,\, \,
(0<\rho<\pi/2),  \nonumber \\
\cos{\tau^{XI}}&=&\sin{\rho^{XI}}(\Omega_{n})^{XI} \, \,,\, \,
\hbox{with} \, \,
0 \le \tau^{XI} \le \pi \,, \nonumber \\
|\overline{\Omega}|^{XI} &=& \left\{
\begin{array}{ll}
         0 & ,\mbox{ if $\alpha \ne 0$}\\
         \frac{2R|\overline{x}|}{R^2+\overline{x}^2}& ,\,
\mbox{if $\alpha = 0$}\end{array}
         \right. \, , \nonumber \\
(\Omega_{n})^{XI} &=& \left\{
        \begin{array}{ll}
         1 &,\, \mbox{ if $\alpha \ne 0$}\\
        \frac{-R^2+\overline{x}^2}{R^2+\overline{x}^2},& \mbox{if $\alpha = 0$}\end{array}
         \right. \,.
\end{eqnarray}

\noindent For region XII we have the same points of the region XI with the
difference that now $-\pi \le \tau^{XII} \le 0$ .

\noindent For region XVI we have that

\begin{eqnarray}
\sec{\rho^{XVI}}&=&\sqrt{1+(\frac{\alpha}{2R})^2+\beta^2} \, \, ,
\nonumber \\
\cos{\tau^{XVI}}&=&\sin{\rho^{XVI}}(\Omega_{n})^{XVI} \, ,\,
\hbox{with} \, \,
0 \le \tau^{XVI} \le \pi \, ,\nonumber \\
(\Omega_{n})^{XVI}&=&\left\{
\begin{array}{ll}\pm
\frac{1}{\sqrt{1+(\frac{2R\beta}{\alpha})^2}}, & \mbox{for
$sgn(\alpha)=\pm 1$ }\\
        0 &, \,\,\mbox{for $\alpha=0$} \end{array}
         \right. \,,
\nonumber \\
 |\overline{\Omega}|^{XVI}&=&
\frac{1}{\sqrt{1+(\frac{\alpha}{2R\beta})^2}} \,.
\end{eqnarray}

\noindent  For region XVII we have the same points of region XVI with the
difference that now $-\pi \le \tau^{XVII} \le 0$ .

\item Regions II, IV, IX, and XIV correspond to the unique point
$\rho=\pi/2, \tau=0, \break |\overline{\Omega}|=0, \Omega_{n}=1$. This corresponds
to point $A$ in figure 2(b).

\item Similarly, regions III, V, VIII, X, XIII and XV correspond
to the unique point  $\rho=\pi/2,
\tau=\pm\pi,|\overline{x}|=0,\Omega_{n}=-1$. This corresponds
to point $B$ in figure 2(b).

\noindent These two special points belong to the boundary of the hyperboloid
and the cutting hyperplane $X_{0}=X_{n}$ because $z\to\infty$.

\end{itemize}

\begin{figure}[!h]
\centerline{\psfig{figure=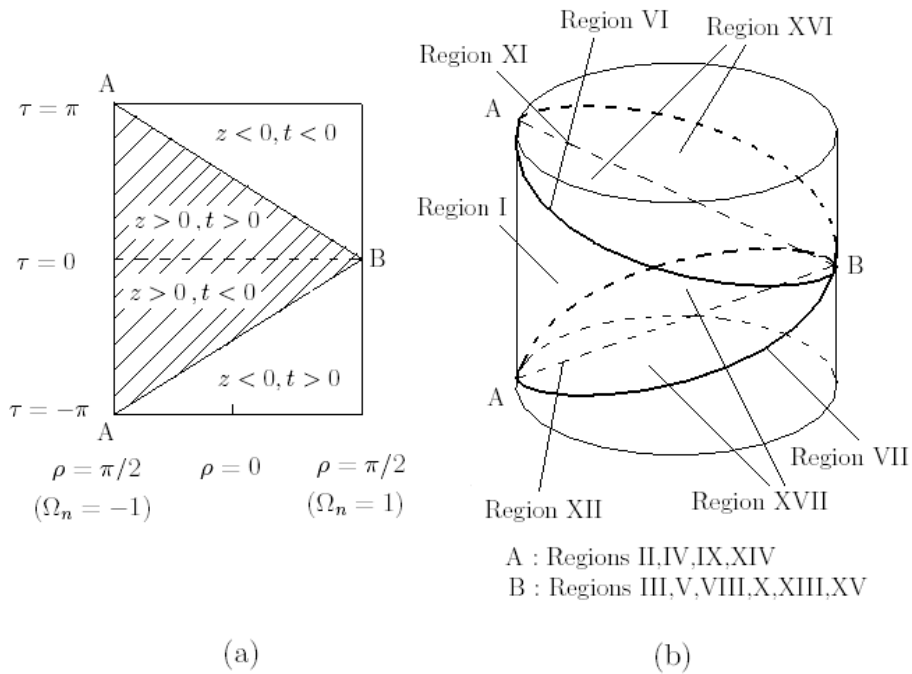,width=12cm}}
\vspace{-1cm}\caption{(a) Penrose diagram for $AdS$ space, the
Poincar\'e $AdS$ space corresponds to the dashed region. (b)
Poincar\'e $AdS$ boundary regions described in the Penrose diagram
in 3D. \label{Fig2}}
\end{figure}

All these results are shown in Figure \ref{Fig2}, where we draw
the Penrose diagram for $AdS$ space.  We can see there that the
Poincar\'e $AdS$ boundary contains points of the global $AdS$
boundary $\rho=\pi/2$ and points of the global $AdS$ bulk. So we
conclude that the Poincar\'e $AdS$ boundary is not the same as the
global $AdS$ boundary.

It is important to observe that regions XI,XII XVI and XVII,  which correspond to points
in the bulk of the global AdS, have a singular metric in the Poincar\'e coordinate system.
These regions can be described in a non singular way returning to global coordinates.

It is also very interesting to note that there are some regions of
the Poincar\'e $AdS$ boundary that corresponds to single points
in global coordinates, so there must be some kind of
identifications between these regions.
Note that these regions may have different embedding space coordinates.
For example, from eq. (\ref{Poinc}) we find that 
region II corresponds to finite  $X_i $ while region 
IX corresponds to  $X_i \,=\,0\,$. This happens because the relation
(\ref{glob})  between global coordinates and embedding coordinates 
is one to one just for points of the hyperboloid but not for points
of the boundary $\rho = \pi/2$. This means that different points of 
the infinity of the hyperboloid are mapped to single points in global coordinates.
This situation is similar to what happens when a 2D Minkowski space
time $(x,t)$ is compactified to the interior of a rectangle and   
the spatial infinity $x \to \infty$ is  mapped
to just one point of the rectangle (the same point for all finite values of $t$).

\section{The Euclidean $AdS$ space}

The isometry group of $AdS_{n+1}$ space in the Lorentzian case is $SO(2,n)$ while in the Euclidean case
it is $SO(1,n+1)$. In the AdS/CFT correspondence the conformal group of the boundary theory is isomorphic to the  isometry group of the bulk gravity theory, so the conformal theories are not the same.
We will see that the boundaries of Euclidean and Lorentzian AdS spaces, where the conformal theories 
are defined are different. 

When calculating conformal correlation functions one finds different results taking  Lorentzian or Euclidean signatures for the AdS metric. As discussed in \cite{Balasubramanian:1998de}, in the Lorentzian case there are
normalizable modes in the solutions of the bulk fields. These normalizable modes are related to states in the boundary CFT. So in the Lorentzian case, using the AdS/CFT correspondence one can calculate operator expectation values in excited states of the conformal theory. 

In the Euclidean case there are no normalizable modes in the solutions of the field equations.
The solutions are uniquely determined by the boundary values of the fields. 
In this case one calculates operator expectation values in the vacuum. 

We will now describe the Euclidean AdS boundary in Poincar\'e coordinates. The geometry of this boundary
is more simple than the Lorentzian case.  We can go to Euclidean signature 
performing a rotation in the time coordinate
\begin{equation}
t\to it \, , \label{rotec}
\end{equation}

\noindent This rotation \cite{Aharony} induces
a rotation on the embedding (time-like) coordinate $X_{n+1}$ :

\begin{equation}
X_{n+1}\to iX_{n+1} \, . \label{rotemb}
\end{equation}

So we have that relations (\ref{Poinc}) become
\begin{eqnarray}
X_{0}&=&\frac{1}{2z}(z^2+R^2+{\overline{x}}^2+t^2) \, ,
\nonumber\\
X_{i}&=&\frac{R x^i}{z}\, , \hskip1cm i=1,..,n-1 \,  ,
\nonumber\\
X_{n}&=&-\frac{1}{2z}(z^2-R^2+{\overline{x}}^2+t^2) \, ,
\nonumber\\
X_{n+1}&=&\frac{R t}{z} \, . \label{Poinc2}
\end{eqnarray}

We also see from  (\ref{rotemb}) and  (\ref{emb}) that the
equation of the hyperboloid is now
\begin{equation}
X_{0}^2-X_{n+1}^2-\sum_{i=1}^{n}X_{i}^2 = R^2 \, , \label{eadseq}
\end{equation}

\noindent (note the different sign). This equation defines the
so-called Euclidean $AdS$ space, this equation can be solved by
the following relations :
\begin{eqnarray}
X_{0}&=&R\sec{\rho}\sec{\tau} \, ,
\nonumber\\
X_{i}&=&R\tan{\rho}\Omega_{i}\, , \hskip1cm i=1,..,n \, ,
\nonumber\\
X_{n+1}&=&R\sec{\rho}\tan{\tau} \,  \label{glob2}
\end{eqnarray}

\noindent where $\sum_{i=1}^{n}{{\Omega_{i}}^2}=1$. The relations
(\ref{glob2}) define a new chart that we will call semi-global
chart because  it  can describe only one half of the hyperboloid.
For example, the first half of the hyperboloid ($z>0$)
corresponding to the region $X_{0}\ge R$ can be obtained by taking
in (\ref{glob2}) the range $-\frac{\pi}{2}<\tau<\frac{\pi}{2}$ .
The other half ($z<0$) corresponding to the region $X_{0}\le -R$
can be also obtained by working in the range
$\frac{\pi}{2}<\tau<\frac{3\pi}{2}$.

It is interesting to note that the cutting hyperplane
$X_{0}=X_{n}$, described in section II, does not contain points of
the entire Euclidean $AdS$ space because this condition is not
compatible with (\ref{eadseq}).

We can find a map of the Euclidean $AdS$ boundary to this
semi-global chart. For this purpose, we obtain 
relations between Euclidean Poincar\'e coordinates
and semi-global coordinates in the same way as
in section III:

\begin{eqnarray}
\sec{\rho} &=& \frac{1}{2R
|z|}\sqrt{(z^2+R^2+{\overline{x}}^2+t^2)^2
- (2R t )^2 } \, , \nonumber\\
\sec{\tau}&=&sgn(z)
\frac{z^2+R^2+{\overline{x}}^2+t^2}{\sqrt{(z^2+R^2+{\overline{x}}^2+t^2)^2-(2R
t)^2}} \, ,\nonumber \\
\tan{\tau}&=&sgn(z)\frac{2R
t}{\sqrt{(z^2+R^2+{\overline{x}}^2+t^2)^2-(2R
t)^2}} \, , \nonumber \\
 |\overline{\Omega}| &=& \frac{2R
|\overline{x}|}{\sqrt{(z^2+R^2+{\overline{x}}^2+t^2)^2 - (2R
t)^2-(2R z)^2}} \, , \,  \, \hbox{where} \, |\overline{x}| \equiv
\sqrt{\overline{x}^2} \, \, , \nonumber  \\
\Omega_{n}&=&sgn(z)\frac{z^2-R^2+{\overline{x}}^2+t^2}{\sqrt{(z^2+R^2+
{\overline{x}}^2+t^2)^2
- (2R t)^2 - (2R z )^2}} \, . \nonumber \\
\end{eqnarray}

Region I, defined in the last section, corresponds in
Euclidean Poincar\'e coordinates to an Euclidean space and can be
mapped to the semi-global chart. The semi-global coordinates
corresponding to this region are given by

\begin{eqnarray}
\rho^{I} &=& \pi/2 \, , \nonumber \\
\sec{\tau^{I}}&=&\frac{R^2+{\overline{x}}^2+t^2}{\sqrt{(R^2+{\overline{x}}^2+t^2)^2-(2R
t)^2}} \,  , \nonumber \\
 \tan{\tau^{I}}&=&\frac{2R
t}{\sqrt{(z^2+R^2+{\overline{x}}^2+t^2)^2-(2R t)^2}} \, , \nonumber \\
|\overline{\Omega}|^{I}&=& \frac{2R
|\overline{x}|}{\sqrt{(R^2+{\overline{x}}^2+t^2)^2 - (2R t)^2}} \,
, \nonumber \\
(\Omega_{n})^{I}&=&\frac{-R^2+{\overline{x}}^2+t^2}{\sqrt{(R^2+{\overline{x}}^2+t^2)^2
- (2R t)^2 }} \, .
\end{eqnarray}

The other regions  correspond in semi-global coordinates to the
unique point $\rho = \pi/2, \, \break \tau=0\, \, |\overline{\Omega}|=0
\, , \, \Omega_{n}=1 $.  This point does not belong to region I,
it belongs to the boundary of the hyperboloid and to the
 cutting hyperplane $X_{0}=X_{n}$. We conclude that the Euclidean AdS boundary in Poincar\'e coordinates
is the Euclidean space region $\R^n \,$ at $z=0$ plus one point at infinity. This is different
from the Lorentzian case where the boundary is $\R^{1,n-1}\,$ plus the other regions described in section 
 III.

This way we found an equivalence between
 the Euclidean Poincar\'e chart and the semi-global chart,
 because they describe in fact the same space with the same boundary.

\section{ Conclusions}
 
We found a detailed map of the boundary of Poincar\'e $AdS$ space
to the global coordinate chart. We studied also the origin of 
Poincar\'e $AdS$ space as a subspace of the global $AdS$.  
From the map, described in section III,
we conclude that if a Poincar\'e $AdS$ space corresponds in fact to a
subspace of the global $AdS$ we have to impose the identification
of some regions of  its boundary. In the Euclidean case, we proposed 
a semi-global chart that describes the same space as
the Euclidean Poincar\'e chart. The identification of boundary regions 
of the Euclidean Poincar\'e chart is trivial. 

The AdS/CFT correspondence is usually interpreted as a realization of the holographic principle.
This interpretation depends on the assumption that the region $z= 0$ plus some isolated points at infinity
is the boundary of the Poincar\'e AdS space. Since the isolated points do not carry degrees of freedom,
the correspondence between a field theory in the Poincar\'e AdS and a theory on $z=0 $ is taken as a
bulk/boundary correspondence.
Here we conclude that this interpretation is not correct
because the Minkowski space, even after compactification, is not the complete boundary of the
Poincar\' e AdS space. This boundary actually includes other 
regions of the global AdS space (see for instance regions XVI, XVII). 
The bulk/boundary interpretation is correct for the Euclidean case, where indeed the boundary is the region $z = 0$ plus a point at infinity.

\acknowledgements The authors are partially
supported by CNPq, CAPES ,CLAF and FAPERJ.

\end{document}